\documentclass[conference]{IEEEtran}
\usepackage{cite}
\usepackage{amsmath,amssymb,amsfonts}
\usepackage{algorithm}
\usepackage[algo2e]{algorithm2e}
\usepackage{algpseudocode}
\usepackage{graphicx}
\usepackage{textcomp}
\usepackage{xcolor}	
\usepackage[caption=false,font=footnotesize]{subfig}
\usepackage{stfloats}
\usepackage{epstopdf}
\usepackage{acronym}
\usepackage[bookmarksopen=true]{hyperref}
\usepackage{mathrsfs}
\usepackage{float}
\usepackage{placeins}
\usepackage{epstopdf}
\usepackage{subfloat}

\usepackage{flushend} 
\usepackage{lipsum}

\def\BibTeX{{\rm B\kern-.05em{\sc i\kern-.025em b}\kern-.08em
		T\kern-.1667em\lower.7ex\hbox{E}\kern-.125emX}}
	
\usepackage{amsthm}

\newtheoremstyle{customdef} 
{}                          
{}                          
{\itshape}                         
{}                         
{\bfseries}                 
{.}                         
{ }                         
{}  

\theoremstyle{customdef}
\newtheorem{definition}{Definition} 

\newtheoremstyle{customremark} 
{}                          
{}                          
{}                         
{}                         
{\bfseries}                 
{.}                         
{ }                         
{}                          

\theoremstyle{customremark}
\newtheorem{remark}{Remark}

\newtheoremstyle{customexample} 
{}                          
{}                          
{}                         
{}                         
{\bfseries}                 
{.}                         
{ }                         
{}                          

\theoremstyle{customexample}
\newtheorem{example}{Example}

\newcommand{\trans}{^{\mathrm{T}}}
\newcommand{\herm}{^{\mathrm{H}}}
\newcommand{\vecsym}[1]{\boldsymbol{\rm{#1}}}
\newcommand{\diag}{\mathrm{diag}}

\def\e{\mathrm e}

\def\R{\mathbb{R}}
\def\C{\mathbb{C}}
\def\bField{\mathbb{F}_2}
\def\zeroPair{\mathscr{Z}}
\def\Code{\mathscr{C}}

\def\vector{\vecsym{v}}

\def\complexnormal{\mathcal{CN}}

\def\bMessage{\vecsym{b}}

\def\txSeq{\vecsym{x}}
\def\rxSeq{\vecsym{y}}
\def\noiseSeq{\vecsym{w}}
\def\rxSeqMod{\tilde{\rxSeq}}
\def\rxSeqHat{\hat{\rxSeq}}
\def\modSeq{\vecsym{m}}

\def\bAlpha{\boldsymbol{\alpha}}

\def\numTaps{L}

\def\channelIndex{l_\mathrm{e}}
\def\totIndex{l_\mathrm{t}}
\def\filter{\vecsym{h}}

\def\totalLength{L_{\mathrm{t}}}
\def\effTaps{L_{\mathrm{e}}}

\def\messageBit{b}
\def\numBits{B}
\def\zeroMag{r}
\def\zeroPhase{\psi}
\def\zeroIndex{k}
\def\numZeros{K}
\def\zero{\alpha}

\def\X{X}
\def\Y{Y}
\def\Ytilde{\tilde{\Y}}
\def\Yhat{\hat{\Y}}
\def\yhatelem{\hat{y}}
\def\H{H}
\def\W{W}
\def\channelZero{\beta}
\def\noiseZero{\gamma}
\def\A{A}
\def\T{T}
\def\autoIndex{\ell}
\def\radius{R}
\def\asymFactor{\zeta}

\def\cfo{\Delta f}

\def\normCFO{\xi}
\def\po{\phi}
\def\poEst{\hat{\po}}
\def\bandwidth{W}
\def\baseAngle{\theta_\numZeros}
\def\intCFO{u}
\def\fracCFO{\epsilon}
\def\poMin{\po_\mathrm{min}}
\def\poMax{\po_\mathrm{max}}
\def\numIterations{n_\mathrm{iterations}}
\def\windowSize{\delta}
\def\iterationIndex{\iota}

\def\dftSize{N}
\def\dftIndex{n}
\def\rowIndex{m}
\def\dftIndexHat{\hat{\dftIndex}}
\def\tN{\vecsym{t}_\dftSize}
\def\YNtilde{\tilde{\vecsym{Y}}_\dftSize}
\def\modMat{\vecsym{M}}
\def\modMaty{\tilde{\vecsym{M}}}

\def\EbNo{E_\mathrm{b}/N_0}
\acrodef{WSN}{wireless sensor network}
\acrodef{USRP}{universal software radio peripheral}
\acrodef{SN}{sensor node}
\acrodef{FC}{fusion center}
\acrodef{MAC}{multiple-access channel}
\acrodef{FL}{federated learning}
\acrodef{ED}{edge device}
\acrodef{CS}{compressed sensing}
\acrodef{ES}[BS]{base station}
\acrodef{DCN}{data center network}
\acrodef{RIS}{reconfigurable intelligent surfaces}
\acrodef{IMC}{in-memory computing}
\acrodef{FPGA}{field-programmable gate array}
\acrodef{SDR}{software-defined radio}
\acrodef{PS}{processing system}
\acrodef{SS}{soft synchronization}
\acrodef{IQ}{in-phase/quadrature}
\acrodef{IP}{intellectual property}
\acrodef{DMA}{direct-memory access}
\acrodef{RAM}{random access memory}
\acrodef{CC}{companion computer}
\acrodef{FEE}{function estimation error}
\acrodef{MSK}{minimum-shift keying}
\acrodef{TDMA}{time-domain multiple access}
\acrodef{PLNC}{physical-layer network coding}
\acrodef{UAV}{unmanned aerial vehicle}
\acrodef{LoRa}{Long-Range}
\acrodef{DC}{direct-current}
\acrodef{DAC}{digital-to-analog converter}
\acrodef{ADC}{anlog-to-digital converter}
\acrodef{CS}{complementary sequence}
\acrodef{GCP}{Golay complementary pair}
\acrodef{ANF}{algebraic normal form}
\acrodef{AACF}{aperiodic auto-correlation function}
\acrodef{AACFs}{aperiodic auto-correlation functions}
\acrodef{RM}{Reed-Muller}
\acrodef{MOCZ}{Modulation on conjugate-reciprocal zeros}
\acrodef{MOZ}{Modulation on zeros}
\acrodef{BMOCZ}{binary modulation on conjugate-reciprocal zeros}
\acrodef{J-BMOCZ}[J-BMOCZ]{jutted BMOCZ}
\acrodef{JBMOCZ}[J-BMOCZ]{jutted binary modulation on conjugate-reciprocal zeros}
\acrodef{dizet}[DiZeT]{direct zero-testing}

\acrodef{PUCCH}{physical uplink control channel}
\acrodef{PRACH}{physical random access channel}

\acrodef{OBO}{output-power back-off}
\acrodef{ACLR}{adjacent-channel-leakage ratio}

\acrodef{LDPC}{low-density parity check}

\acrodef{PDF}{probability density function}
\acrodef{CDF}{cumulative distribution function}
\acrodef{CCDF}{complementary cumulative distribution function}

\acrodef{TBMA}{type-based multiple access}

\acrodef{MSFE}{mean-squared function error}
\acrodef{FEE}{function-estimation error}
\acrodef{CER}{computation error rate}
\acrodef{BCER}{block-computation error rate}
\acrodef{CFO}{carrier frequency offset}
\acrodef{TO}{timing offset}
\acrodef{PO}{phase offset}
\acrodef{RSSI}{received signal strength  information}

\acrodef{STLC}{space-time line code}
\acrodef{CCI}{co-channel interference}
\acrodef{CSIT}[CSIT]{\ac{CSI} at the transmitter}
\acrodef{CSIR}[CSIR]{\ac{CSI} at the receiver}
\acrodef{MIMO}{multiple-input-multiple-output}
\acrodef{PC}{phase correction}
\acrodef{ZF}{zero-forcing}
\acrodef{ANOVA}{analysis of variance}

\acrodef{PCA}{principal component analysis}
\acrodef{TIG}{Technical Interest Group}

\acrodef{FSK}{frequency-shift keying}
\acrodef{PPM}{pulse-position modulation}
\acrodef{PAM}{pulse-amplitude modulation}

\acrodef{MRC}{maximum-ratio combining}
\acrodef{HP}{hard-coded participation}
\acrodef{HPA}{hard-coded participation with absentees}
\acrodef{SP}{soft-coded participation}
\acrodef{FSK-MV}{\ac{FSK}-based \ac{MV}}
\acrodef{RF}{radio-frequency}
\acrodef{MF}{matched filter}
\acrodef{PPM}{pulse-position modulation}
\acrodef{CSK}{chirp-shift keying}
\acrodef{PPM-MV}[PPM-MV]{\ac{PPM}-based \ac{MV}}
\acrodef{DFT-s-OFDM}{discrete Fourier transform-spread orthogonal frequency division multiplexing}
\acrodef{SC}{single-carrier}
\acrodef{SGD}{stochastic gradient descent}
\acrodef{signSGD}{sign stochastic gradient descent}

\acrodef{SL}{split learning}
\acrodef{SNR}{signal-to-noise ratio}
\acrodef{RMSE}{root-mean-squared error}
\acrodef{OFDM}{orthogonal frequency division multiplexing}
\acrodef{DFT}{discrete Fourier transform}
\acrodef{PSK}{phase-shift keying}
\acrodef{QAM}{quadrature amplitude modulation}
\acrodef{QPSK}{quadrature phase-shift keying}
\acrodef{PMEPR}{peak-to-mean envelope power ratio}
\acrodef{BER}{bit error rate}
\acrodef{SNR}{signal-to-noise ratio}
\acrodef{PSD}{power spectral density}
\acrodef{SE}{spectral efficiency}
\acrodef{CP}{cyclic prefix}
\acrodef{AWGN}{additive white Gaussian noise}
\acrodef{CFR}{channel frequency response}
\acrodef{CIR}{channel impulse response}
\acrodef{MMSE}{minimum mean-squared error}
\acrodef{LMMSE}{linear minimum mean-squared error}
\acrodef{BPSK}{binary phase shift keying}
\acrodef{BPSK}{quadrature phase shift keying}
\acrodef{BLER}{block error rate}
 \acrodef{ML}{maximum likelihood}
\acrodef{PHY}{physical layer}
\acrodef{PA}{power amplifier}
\acrodef{IDFT}{inverse discrete Fourier transform}
\acrodef{DoF}{degrees-of-freedom}
\acrodef{IoT}{Internet of Things}
\acrodef{mMTC}{massive machine-type communication}
\acrodef{URLLC}{ultra-reliable low-latency communication}
\acrodef{FDE}{frequency-domain equalization}
\acrodef{RF}{radio-frequency}
\acrodef{IM}{index modulation}
\acrodef{MF}{matched filter}
\acrodef{PPM}{pulse-position modulation}

\acrodef{MSE}{mean-squared error}
\acrodef{MRT}{maximum-ratio transmission}
\acrodef{ERC}{equal-ratio combining}
\acrodef{BAA}{broadband analog aggregation}
\acrodef{OBDA}{one-bit broadband digital aggregation}
\acrodef{FEEL}{federated edge learning}
\acrodef{FL}{federated learning}
\acrodef{UL}{uplink}
\acrodef{OAC}{over-the-air computation}
\acrodef{TCI}{truncated-channel inversion}
\acrodef{MV}{majority vote}
\acrodef{CNN}{convolution neural network}
\acrodef{ReLU}{rectified-linear unit}
\acrodef{CSI}{channel state information}
\acrodef{PAPR}{peak-to-average power ratio}
\acrodef{SC}{single-carrier}
\acrodef{iid}[IID]{independent and identically distributed}
\acrodef{RMS}{root-mean-square}
\acrodef{4G}{fourth generation}
\acrodef{5G}{Fifth Generation}
\acrodef{6G}{Sixth Generation}
\acrodef{NR}{New Radio}
\acrodef{LTE}{Long-Term Evolution}
\acrodef{OFDMA}{orthogonal frequency division multiple access}
\acrodef{HARQ}{hybrid automatic repeat request}
\acrodef{D2D}{Device-to-Device}
\acrodef{NOMA}{non-orthogonal multiple access}
\acrodef{OMA}{orthogonal multiple access}
\acrodef{IMT}{International Mobile Telecommunications}
\acrodef{ITU}{International Telecommunication Union}

\acrodef{PDP}{power-delay profile}
\acrodef{TBMA}{type-based multiple access}
\acrodef{ISI}{intersymbol interference}
\acrodef{MLSE}{maximum likelihood sequence estimator}
\acrodef{LTI}{linear time-invariant}
\acrodef{ISAC}{integrated sensing and communication}
\acrodef{DL}{deep learning}
\acrodef{AE}{autoencoder}
\acrodef{AEs}{autoencoders}
\acrodef{MLP}{multi-layer perceptron}
\acrodef{CPC}{cyclically permutable code}
\acrodef{ACPC}{affine cyclically permutable code}
\acrodef{ICFOE}{iterative carrier frequency offset estimator}
\acrodef{BCH}{Bose-Chaudhuri-Hocquenghem}

\begin{document}
	
	\title{Fourier-Domain CFO Estimation Using Jutted Binary Modulation on Conjugate-Reciprocal Zeros}
	
    \author{Parker Huggins, Anthony Joseph Perre, and Alphan \c{S}ahin\\
	Department of Electrical Engineering, University South Carolina, Columbia, SC, USA\\
	Email: \{parkerkh, aperre\}@email.sc.edu, asahin@mailbox.sc.edu 
	}
	\maketitle
	
    \begin{abstract}  	
		In this work, we propose \ac{JBMOCZ} for non-coherent communication under a \ac{CFO}. By introducing asymmetry to the Huffman BMOCZ zero constellation, we exploit the identical aperiodic auto-correlation function of BMOCZ sequences to derive a Fourier-domain metric for \ac{CFO} estimation. Unlike the existing methods for Huffman BMOCZ, which require a \ac{CPC} for pilot-free \ac{CFO} correction, \ac{JBMOCZ} enables the estimation of a \ac{CFO} without the use of pilots or channel coding. Through numerical simulations in additive white Gaussian noise and fading channels, we show that the \ac{BER} loss of \ac{JBMOCZ} under a \ac{CFO} is just 1~dB over Huffman BMOCZ without a \ac{CFO}. Furthermore, the results show that coded \ac{JBMOCZ} achieves better \ac{BER} performance than Huffman BMOCZ with a \ac{CPC}. 
	\end{abstract}

	\begin{IEEEkeywords}
		BMOCZ, \ac{CFO}, modulation, polynomial, zeros
	\end{IEEEkeywords}

	\acresetall
	\section{Introduction} \label{sec:intro}
	
	\ac{MOZ} is an emerging non-coherent communication scheme with several attractive characteristics. First proposed in~\cite{walk2017short}, the principle of \ac{MOZ} is to encode digital information into the zeros (i.e., roots) of the baseband signal's $z$-transform, which is a polynomial of finite degree. In this way, the impact of the channel becomes polynomial multiplication in the $z$-domain, which preserves the zeros of the transmitted polynomial, regardless of the \ac{CIR} realization. Since the zero structure of the transmitted polynomial is unaffected after passing through the channel, \ac{MOZ} enables non-coherent detection at the receiver. 

	In~\cite{walk2019principles}, the authors introduce a binary \ac{MOZ} scheme called \ac{BMOCZ}. Using \ac{BMOCZ}, a simple \ac{dizet} decoder can be implemented at the receiver, where the transmitted bits are estimated by evaluating the received polynomial at each zero in the constellation~\cite{walk2019principles,walk2020practical}. The performance of the \ac{dizet} decoder is sensitive to the zero constellation, since the placement of the zeros in the complex plane controls their stability under additive noise perturbing the coefficients~\cite{Wilkinson1984}. Presently, the leading solution for zero stability is Huffman \ac{BMOCZ}~\cite{walk2019principles}, where the zeros are placed uniformly along two concentric circles. With this approach, the coefficients of each transmit polynomial form a Huffman sequence, which have an impulse-like \ac{AACF}. Huffman \ac{BMOCZ} has been the subject of many recent studies, including works on integrated sensing and communication~\cite{dehkordi2023integrated}, over-the-air computation~\cite{csahin2024over}, and novel $z$-domain multiplexing techniques~\cite{eren2025non}.
	
	Of particular interest in this work, however, is the resiliency of \ac{BMOCZ} against a \ac{CFO}, which is a ubiquitous hardware impairment in communication systems. The topic of a \ac{CFO} is addressed for Huffman \ac{BMOCZ} in~\cite{walk2020practical}, where a two-step correction procedure is proposed. In the first step, an oversampled \ac{dizet} decoder is used to correct the fractional component of the \ac{CFO}, yielding to zeros which correspond to a cyclic permutation of the transmitted message. To estimate the integer part of the \ac{CFO} in a second step, the authors utilize an \ac{ACPC}, where all cyclic permutations of a valid codeword correspond to the same message. Although effective for \ac{CFO} correction, in practice the \ac{ACPC} has several drawbacks. In particular, it fixes the form of channel coding and places numerous constraints on the message length and coding rate. In turn, this limits the flexibility of \ac{BMOCZ}, preventing its implementation with more commonly used codes, such as polar and \ac{LDPC} codes.
	
	In this study, we propose an asymmetric \ac{BMOCZ} zero constellation, called \ac{J-BMOCZ}, that enables the receiver to estimate a \ac{CFO} using the \ac{IDFT}. Our approach leverages the common \ac{AACF} of \ac{BMOCZ} sequences to achieve \ac{CFO} estimation without pilots or channel coding. We demonstrate the efficacy of \ac{J-BMOCZ} under a \ac{CFO} using both \ac{BER} and \ac{BLER} simulations in \ac{AWGN} and fading channels. Furthermore, we show that the \ac{BER} performance of coded \ac{J-BMOCZ} outperforms Huffman \ac{BMOCZ} with an \ac{ACPC}.
	
	\emph{Organization:} The remainder of the paper is organized as follows: Section~\ref{sec:model} describes the system model, including a review of \ac{BMOCZ} and zero rotation; Section~\ref{sec:methods} introduces the proposed \ac{CFO} estimation algorithm together with \ac{J-BMOCZ}; Section~\ref{sec:results} presents the results of numerical simulations; and Section~\ref{sec:conclusion} concludes the paper. 
	
	\emph{Notation}: The set of real and complex numbers are denoted by $\R$ and $\C$, respectively, and the binary field of dimension $n$ is denoted by $\bField^n$. We use $[N]=\{0,1,\hdots,N-1\}$ to denote the set of the first $N$ non-negative integers. The complex conjugate of $z\in\C$ is expressed as $\overline{z}$. Small boldface letters denote vectors, while large boldface letters denote matrices. We write the $\ell_2$-norm of $\vector\in\C^N$ as $||\vector||_2=\sqrt{\vector\herm\vector}$, and the Hadamard product is denoted by $\odot$. We write the complex normal distribution with mean zero and variance $\sigma^2$ as $\complexnormal(0,
	\sigma^2)$.
	
	\section{System Model} \label{sec:model}
	
	\subsection{Preliminaries on \ac{BMOCZ}} \label{subsec:preliminaries}
	
	The principle of \ac{BMOCZ} is to modulate information onto the zeros of the baseband signal's $z$-transform. Considering a $\numZeros$-bit binary message $\bMessage=(b_0,b_1,\hdots,b_{\numZeros-1})\in\bField^\numZeros$, the $\zeroIndex$th message bit is mapped to the zero of a polynomial as 
	\begin{equation} \label{eq:zero_mapping}
		\zero_\zeroIndex \triangleq
		\begin{cases}
			\zeroMag_\zeroIndex\e^{j\zeroPhase_\zeroIndex}, & \messageBit_\zeroIndex=1\\
			\zeroMag_\zeroIndex^{-1}\e^{j\zeroPhase_\zeroIndex}, & \messageBit_\zeroIndex=0
		\end{cases}\:, \ \ \zeroIndex\in[K]\:,
	\end{equation} 
	where $\zeroMag_\zeroIndex>1$ and $\zeroPhase_\zeroIndex\in[0,2\pi)$ define the magnitude and phase of the $\zeroIndex$th zero, respectively. 
	The zeros in~\eqref{eq:zero_mapping} are called \emph{conjugate-reciprocal} zeros, for $\overline{(\zeroMag_\zeroIndex\e^{j\zeroPhase_\zeroIndex})^{-1}}=\zeroMag_\zeroIndex^{-1}\e^{j\zeroPhase_\zeroIndex}$. With~\eqref{eq:zero_mapping}, each binary message $\bMessage\in\bField^\numZeros$ maps to a distinct zero pattern $\bAlpha=[\zero_0,\zero_1,\hdots,\zero_{\numZeros-1}]\trans\in\C^\numZeros$, which, by the fundamental theorem of algebra, uniquely defines the polynomial
	\begin{equation} \label{eq:zeros_to_poly}
		\X(z)=\sum_{\zeroIndex=0}^{\numZeros}x_\zeroIndex z^\zeroIndex=x_\numZeros\prod_{\zeroIndex=0}^{\numZeros-1}(z-
		\zero_\zeroIndex)\:,
	\end{equation}
	up to the complex scalar $x_\numZeros\neq0$. The polynomial $\X(z)$ has $\numZeros+1$ coefficients and is the $z$-transform of the sequence $\txSeq=[x_0,x_1,\hdots,x_\numZeros]\trans\in\C^{\numZeros+1}$. To render the mapping in~\eqref{eq:zeros_to_poly} one-to-one, the leading coefficient $x_\numZeros$ is selected such that the sequence energy is normalized to $\numZeros+1$, i.e., $||\txSeq||_2^2=\numZeros+1$. For a given $\numZeros$, the set of all normalized polynomial sequences $\txSeq$ defines a codebook, which is denoted by $\Code^\numZeros$. 
	
	Assuming that the channel is \ac{LTI} with an impulse response $\filter=[h_0,h_1,\hdots,h_{\numTaps-1}]\trans\in\C^{\effTaps}$ of effective length $\effTaps\geq1$, the received sequence is given by $\rxSeq=\txSeq\ast\filter+\noiseSeq\in\C^{\totalLength}$, where $\totalLength=\numZeros+\effTaps$ and $\noiseSeq=[w_0,w_1,\hdots,w_{\totalLength-1}]\trans\in\C^{\totalLength}$ is a sequence of \ac{AWGN}. Invoking the convolution theorem, the input-output relation of the channel can be recast in the $z$-domain as
	\begin{equation} \label{eq:received_poly}
		\Y(z)=\X(z)\H(z)+\W(z)\:,
	\end{equation}
	where $\H(z)$ and $\W(z)$ are the $z$-transforms of $\filter$ and $\noiseSeq$, respectively. Since $\H(z)$ and $\W(z)$ are both polynomials in $z\in\C$, they can be factored as $\H(z)=h_{\effTaps-1}\prod_{\channelIndex=0}^{\effTaps-2}(z-\channelZero_{\channelIndex})$ and $\W(z)=w_{\totalLength-1}\prod_{\totIndex=0}^{\totalLength-2}(z-\noiseZero_{\totIndex})$. Therefore, although perturbed by noise, the $\numZeros$ data zeros $\{\zero_\zeroIndex\}$ are roots of the received polynomial in~\eqref{eq:received_poly}, in addition to the $\effTaps-1$ zeros $\{\channelZero_{\channelIndex}\}$ introduced by the channel. In this work, we assume that $\effTaps=1$, i.e., the channel is flat-fading with a complex channel coefficient $h\sim\complexnormal(0,1)$, and no zeros in $\Y(z)$ are introduced by the channel. In practice, this can be achieved by integrating \ac{BMOCZ} within an \ac{OFDM} framework and appropriately mapping the polynomial coefficients to time-frequency resources, as demonstrated in~\cite{huggins2024optimal}. 
	
	We now define the \ac{AACF} of a general sequence $\txSeq\in\C^{\numZeros+1}$, which will be important to us moving forward.
	\begin{definition} \label{def:autoCorr}
		The AACF of $\txSeq=[x_0,x_1,\hdots,x_\numZeros]\trans\in\C^{\numZeros+1}$ is the complex-valued function $\A(z) \triangleq \sum_{\autoIndex=-\numZeros}^{\numZeros}a_\autoIndex z^\ell$, where $a_\autoIndex$ is defined as 
		\begin{equation}
			a_\autoIndex \triangleq
			\begin{cases}
				\sum_{i=0}^{\numZeros-\autoIndex}\overline{x_i}x_{i+\autoIndex}, & 0\leq\autoIndex\leq\numZeros\\
				\sum_{i=0}^{\numZeros+\autoIndex}x_i\overline{x_{i-\autoIndex}}, & -\numZeros\leq\autoIndex<0\\
				0, & \textup{otherwise}
			\end{cases}\:.
		\end{equation}
	\end{definition}
	\noindent
	For \ac{BMOCZ}, the \ac{AACF} $\A(z)$ can be directly calculated from the zeros of $\X(z)$ as
	\begin{align} \label{eq:auto_poly}
		\A(z)&=\sum_{\autoIndex=-\numZeros}^{\numZeros}a_\autoIndex z^\ell=\X(z)\overline{\X(1/\overline{z})} \notag\\
		&=z^{-\numZeros}x_\numZeros \overline{x_0} \prod_{\zeroIndex=0}^{\numZeros-1}(z-\zero_k)\prod_{\zeroIndex=0}^{\numZeros-1}(z-1/\overline{\zero_\zeroIndex})\:.
	\end{align}
	A key observation is that the \ac{AACF} in~\eqref{eq:auto_poly} is a function of all the roots defined by~\eqref{eq:zero_mapping}. This elucidates an important property of \ac{BMOCZ}, which we exploit in Section~\ref{subsec:fourier_approach}: all sequences $\txSeq\in\Code^\numZeros$ have an \emph{identical} \ac{AACF}. 

	Drawing on Huffman's work in~\cite{huffman1962generation}, which characterizes the zero locations of impulse-equivalent \ac{AACF}s, the authors of~\cite{walk2019principles} introduce \emph{Huffman \ac{BMOCZ}}, where the zero pairs in~\eqref{eq:zero_mapping} are placed uniformly along two concentric circles in the complex plane, i.e., $\zeroMag_\zeroIndex\triangleq\radius>1$ and $\zeroPhase_\zeroIndex\triangleq2\pi\zeroIndex/\numZeros$. Using this design, a simple decoder is derived that evaluates the polynomial in~\eqref{eq:received_poly} at the zeros in each conjugate-reciprocal zero pair $\zeroPair_k\triangleq\{\zero_k,1/\overline{\zero_k}\}$ --- this is the \ac{dizet} decoding rule:
	\begin{equation} \label{eq:dizet}
		\hat{b}_k =
		\begin{cases}
			1, & |\Y(\zeroMag_\zeroIndex\e^{j\zeroPhase_\zeroIndex})| < \zeroMag_\zeroIndex^\numZeros|\Y(	\zeroMag_\zeroIndex^{-1}\e^{j\zeroPhase_\zeroIndex})|\\
			0, & \text{otherwise}
		\end{cases}\:, \ \ \zeroIndex\in[\numZeros]\:.
	\end{equation}
	The evaluation $|\Y(\zeroMag_\zeroIndex^{-1}\e^{j\zeroPhase_\zeroIndex})|$ is scaled by $\zeroMag_\zeroIndex^\numZeros$ to balance the growth of $\Y(z)$ for $|z|>1$ and thereby ensure a fair comparison between the tested zeros. See~\cite[Section~III-D]{walk2019principles} for more details.
	
		
	\subsection{Carrier Frequency Offset and Zero Rotation} \label{subsec:cfo}
	
\begin{figure*}[t]
		\centering
		
		\subfloat[Zero constellation.]{\includegraphics[width=1.75in]{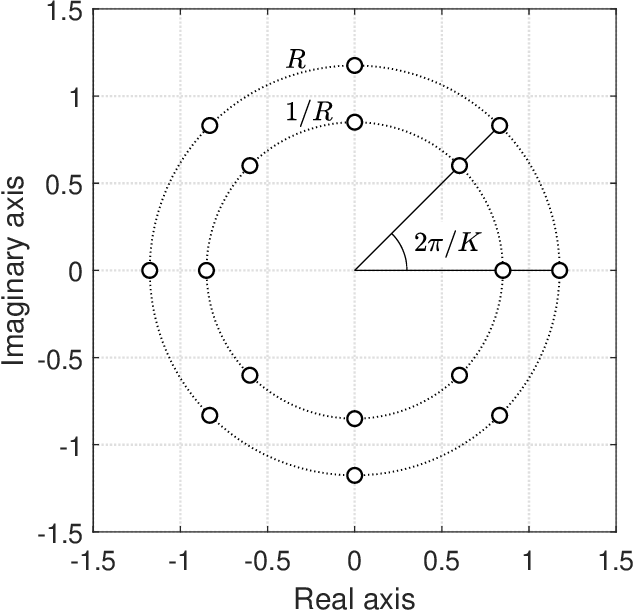}\label{subfig:huffman_constellation}}~
		\subfloat[Transmitted zeros.]{\includegraphics[width=1.75in]{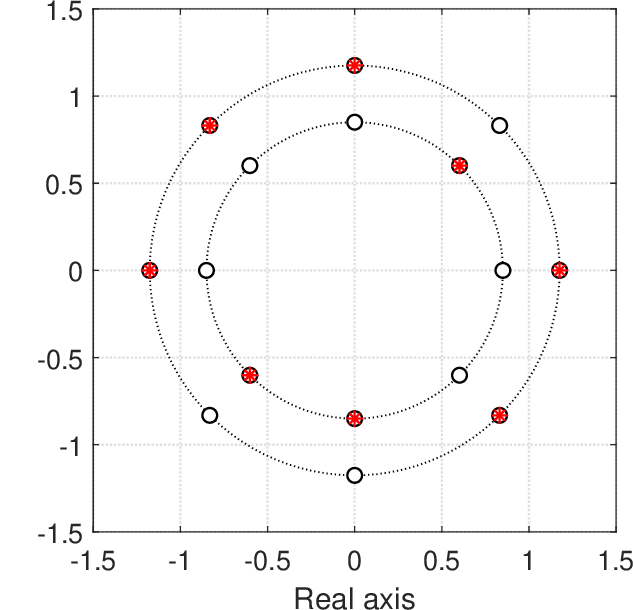}\label{subfig:huffman_tx}}~
		\subfloat[Received zeros.]{\includegraphics[width=1.75in]{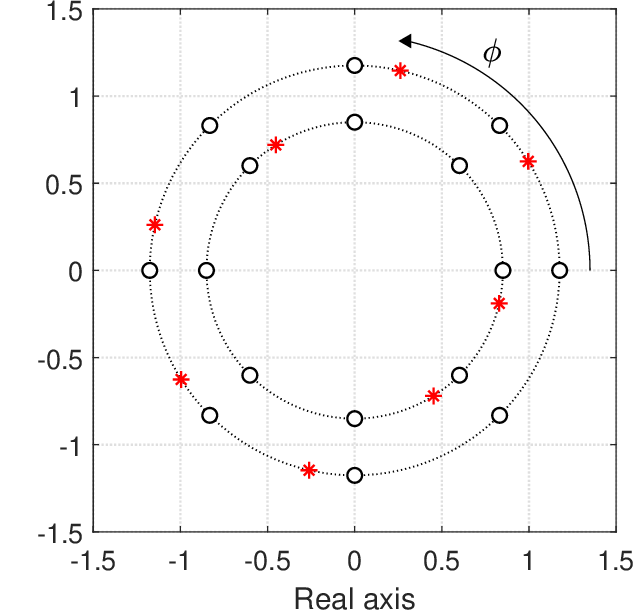}\label{subfig:huffman_rx}}~
		\subfloat[Corrected zeros.]{\includegraphics[width=1.75in]{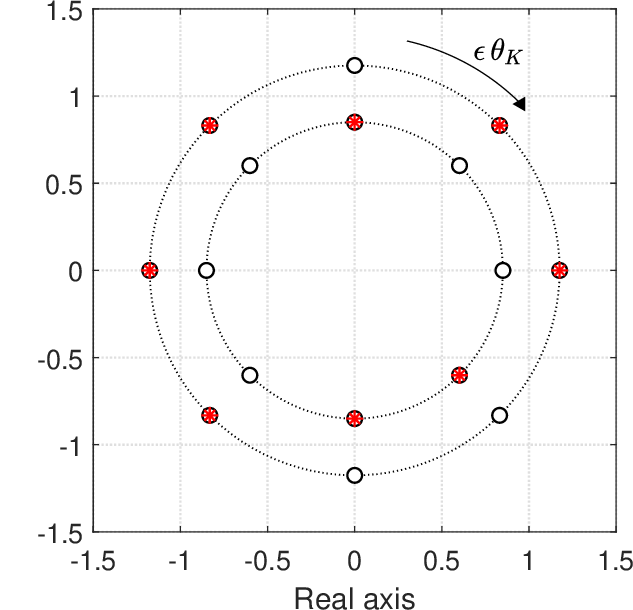}\label{subfig:huffman_corrected}}
			
		\caption{Example Huffman \ac{BMOCZ} zero pattern for $\numZeros=8$ and $\radius=1.176$. (a) Full zero constellation with $2\numZeros$ zero positions. (b) Transmitted zeros corresponding to the message $\bMessage=(1,0,1,1,1,0,0,1)$. (c) Received zeros rotated by $\po=(12/7)\baseAngle$ radians due to a \ac{CFO}. (d) Received zeros after correcting $\fracCFO$, which correspond to a cyclic permutation of the transmitted message, i.e., $\hat{\bMessage}=(1,1,0,1,1,1,0,0)$.}	
		\label{fig:huffman_rotation}
	\end{figure*}

	Consider a \ac{CFO} of $\cfo$~Hz. For a bandwidth $\bandwidth\geq\cfo$, the resulting normalized \ac{CFO}, defined as $\normCFO\triangleq\Delta f/\bandwidth\in[0,1]$, phase modulates the received sequence as
	\begin{align} \label{eq:modulation}
		\rxSeqMod=\modSeq_\po \odot \rxSeq=\modSeq_\po \odot (\txSeq h+\noiseSeq)\:,
	\end{align}
	where $\modSeq_\po=[1,\e^{j\po},\hdots,\e^{j\po\numZeros}]\trans\in\C^{\numZeros+1}$ with $\po=2\pi\normCFO$.\footnote{For consistency with~\cite{walk2020practical}, we consider $\normCFO\in[0,1]$ such that $\po\in[0,2\pi]$. However, in practice a large \ac{CFO} results in \ac{ISI}. A more realistic approach is to bound the \ac{CFO} based on system parameters. For example, in a 20~MHz channel, the 802.11 standard specifies a $\pm25$~ppm oscillator tolerance at 2.4~GHz, corresponding to a maximum \ac{CFO} of $\pm120$~kHz and $\normCFO\in[-0.006,0.006]$. Nevertheless, in this work we assume zero \ac{ISI}.} Excluding noise and transforming~\eqref{eq:modulation} into the $z$-domain, we then obtain
	\begin{align} \label{eq:rotation}
		\Ytilde(z)=\sum_{\zeroIndex=0}^{\numZeros}y_\zeroIndex\e^{j\po\zeroIndex}z^\zeroIndex&=\Y(\e^{j\po}z)\notag\\
		&=\e^{j\po\numZeros}x_\numZeros h\prod_{\zeroIndex=0}^{\numZeros-1}(z-\zero_\zeroIndex\e^{-j\po})\:.
	\end{align}
	Hence, in the presence of a \ac{CFO}, the $\numZeros$ zeros of $\Y(z)$ are rotated through an angle $\po$.\footnote{For \ac{BMOCZ}-based \ac{OFDM} implemented with frequency mapping~\cite{huggins2024optimal}, one may instead consider a timing offset, which induces the same zero rotation we discuss here.} For Huffman \ac{BMOCZ}, the zero constellation is rotationally symmetric such that any estimate of $\po$ is ambiguous modulo $2\pi/\numZeros$. To address this impairment, the authors in~\cite{walk2020practical} decompose $\po$ as $\po=(\intCFO+\fracCFO)\baseAngle$, where $\intCFO\in[\numZeros]$ and $\fracCFO\in[0,1)$ are integer and fractional multiples of the base angle $\baseAngle\triangleq2\pi/\numZeros$, respectively. While the estimation of $\fracCFO$ is straightforward and can be achieved using an oversampled \ac{IDFT}-based \ac{dizet} decoder, the estimation of $\intCFO$ relies upon a \ac{CPC}. Fig.~\ref{fig:huffman_rotation} provides an example Huffman \ac{BMOCZ} zero pattern, which illustrates the zero rotation problem, as well as the fractional \ac{CFO} correction. In the following section, we introduce an approach to correct zero rotation that, together with the proposed constellation of Section~\ref{subsec:constellation}, can directly obtain an estimate of the angle $\po\in[0,2\pi)$ without a \ac{CPC}.  
	
	\section{Methodology} \label{sec:methods}
	
	\subsection{Fourier-Domain \ac{CFO} Estimation} \label{subsec:fourier_approach}
	
	Evaluating $\A(z)$ on the unit circle $\{z=\e^{j\omega} \, | \, \omega\in[0,2\pi)\}$, we obtain
	\begin{equation} \label{eq:psd}
		\A(\e^{j\omega}) = \X(\e^{j\omega})\overline{\X(\e^{j\omega})} = |\X(\e^{j\omega})|^2\:,
	\end{equation}
	which, by recalling that $\A(z)$ is fixed, holds if and only if $|\X(\e^{j\omega})|$ is the same for all $\txSeq\in\Code^\numZeros$. Hence, let us define the \emph{template transform} $\T(\omega)\triangleq|\X(\e^{j\omega})|$, which captures the expected ``shape" of $|\Ytilde(\e^{j\omega})|$. Then, since the zero rotation in~\eqref{eq:rotation} induces a frequency shift in the evaluation $\Y(\e^{j\omega})$, i.e., $\Ytilde(\e^{j\omega})=\Y(\e^{j(\omega+\po)})$, we can estimate $\po$ by cross-correlating $|\Ytilde(\e^{j\omega})|$ with $\T(\omega)$. To that end, we uniformly sample $\T(\omega)$ $\dftSize$ times and define the \emph{template vector} $\tN$ as 
	\begin{equation}
		\tN \triangleq \left[|\X(\e^{j\omega_0})|, |\X(\e^{j\omega_1})|, \hdots, |\X(\e^{j\omega_{\dftSize-1}})|\right]\trans\in\R^{\dftSize}\:,
	\end{equation}
	where $\omega_\dftIndex\triangleq2\pi\dftIndex/\dftSize$ for $\dftIndex\in[\dftSize]$. Note that $\tN$ is equivalent to the magnitude of the $\dftSize$-point \ac{IDFT} of $\txSeq\in\Code^\numZeros$, which is known at the receiver. To identify the rotation $\poEst$ that maximizes the cross-correlation between $|\Ytilde(\e^{j\omega})|$ and $\T(\omega)$, we utilize a modulation matrix to search linearly over rotations in the interval $[\poMin,\poMax)\subseteq[0,2\pi)$ as
		\begin{equation} \label{eq:modMat}
		\modMat \triangleq
		\begin{bmatrix}
			1 & 1 & \hdots & 1\\
			\e^{j\po_0} & \e^{j\po_1} & \hdots & \e^{j\po_{\dftSize-1}}\\
			\vdots & \vdots & \ddots & \vdots\\
			\e^{j\po_0\numZeros} & \e^{j\po_1\numZeros} & \hdots & \e^{j\po_{\dftSize-1}\numZeros}
		\end{bmatrix}\in\C^{(\numZeros+1)\times\dftSize}\:,
	\end{equation}
	where $\po_\dftIndex\triangleq(\poMax-\poMin)\dftIndex/\dftSize+\poMin$ for $\dftIndex\in[\dftSize]$. The columns of $\modMaty=\diag(\rxSeqMod)\modMat$ then comprise $\dftSize$ distinct modulated copies of the received sequence in~\eqref{eq:modulation}, which we transform into the Fourier domain to define $\YNtilde\in\R^{\dftSize\times\dftSize}$, where
	\begin{equation} \label{eq:dftAbsRX}
		(\Ytilde_\dftSize)_{\rowIndex,\dftIndex} = \left|\Ytilde(\e^{j(\omega_\rowIndex-\po_\dftIndex)})\right|\:, \ \ m,n\in[\dftSize]\:.
	\end{equation}
	Here, the $\dftIndex$th column of $\YNtilde$ is simply the magnitude of the $\dftSize$-point \ac{IDFT} of $[\tilde{y}_0,\tilde{y}_1\e^{j\po_\dftIndex},\hdots,\tilde{y}_\numZeros\e^{j\po_\dftIndex\numZeros}]\trans\in\C^{\numZeros+1}$. Thus, to approximate the cross-correlation of $|\Ytilde(\e^{j\omega})|$ with $\T(\omega)$, we compute the inner product of $\tN$ with the columns of $\YNtilde$ and estimate $\po$ as  
	\begin{equation} \label{eq:offsetEst}
		\poEst=(\poMax-\poMin)\frac{\dftIndexHat}{\dftSize} \ \ \text{with} \ \ \dftIndexHat\triangleq\arg\max_{\dftIndex\in[\dftSize]}(\tN\trans\YNtilde)_\dftIndex\:.
	\end{equation}
	After obtaining $\poEst$, the received sequence $\rxSeqMod$ is corrected via modulation with $\modSeq_{-\poEst}=[1,\e^{-j\poEst},\hdots,\e^{-j\poEst\numZeros}]\trans\in\C^{\numZeros+1}$ as $\rxSeqHat=\modSeq_{-\poEst}\odot\rxSeqMod$. The corrected sequence $\rxSeqHat\in\C^{\numZeros+1}$ defines the coefficients of the polynomial $\Yhat(z)$, which is passed to the \ac{dizet} decoder in~\eqref{eq:dizet}.
	
	In~\eqref{eq:offsetEst}, the resolution of $\poEst$ is controlled by $\dftSize$ and can be made arbitrarily fine by choosing $\dftSize\gg\numZeros+1$. However, to avoid the overhead of a large \ac{IDFT} operation in~\eqref{eq:dftAbsRX}, we propose an iterative approach to compute $\poEst$. The approach works by refining an estimate of $\po$ over $\numIterations\geq1$ iterations. In the first iteration, we set $\poMin=0$ and $\poMax=2\pi$ as to quantize the full interval $[0,2\pi)$ into $\dftSize$ uniform bins. By choosing a comparatively small $\dftSize$, this yields a coarse estimate $\poEst$, which we use to update the lower and upper search bounds as $\poMin=\poEst-\windowSize/\iterationIndex$ and $\poMax=\poEst+\windowSize/\iterationIndex$, where $\windowSize\in(0,1)$ is a fixed \emph{window size} and $\iterationIndex=1,2,\hdots,\numIterations$ is the iteration index. Hence, with each consecutive iteration the search interval $[\poMin,\poMax)$ shrinks, allowing $\poEst$ to converge to a fine estimate of the true angle $\po$. A summary of this approach is provided in Algorithm~\ref{alg:ICFOE}.
	
	\begin{remark}
		The presented \ac{CFO} estimation algorithm only applies to \ac{BMOCZ} zero constellations yielding a non-periodic template vector over $\dftIndex\in[\dftSize]$. In particular, the proposed algorithm cannot estimate the \ac{CFO} for Huffman \ac{BMOCZ}, since in this case the template vector is sinusoidal (see Fig.~\ref{fig:template_plot}). In the following section, we introduce a Huffman-like but asymmetric zero constellation for robust \ac{CFO} estimation.
	\end{remark}
	
	\begin{algorithm}[t]
		
		\caption{Iterative CFO Estimation} \label{alg:ICFOE}
		\KwIn{$\rxSeqMod$, $\tN$, $\numZeros$, $\dftSize$, $\windowSize$, $\numIterations$}
		\KwOut{$\poEst$}
		
		Initialize $\poMin=0$ and $\poMax=2\pi$
		
		\For{$\iterationIndex=1$ $\mathrm{to}$ $\numIterations$}
		{

			$\modMat\gets$\eqref{eq:modMat} \hfill (generate modulation matrix)
			
			$\modMaty\gets\diag(\rxSeqMod)\modMat$ \hfill (modulate received sequence)
			
			$\YNtilde\gets$\eqref{eq:dftAbsRX} \hfill (oversample and take \ac{IDFT})
			
			$\poEst\gets$\eqref{eq:offsetEst} \hfill (estimate \ac{CFO})
			
			$\poMin\gets\poEst-\windowSize/\iterationIndex$ \hfill (update lower search bound)
			
			$\poMax\gets\poEst+\windowSize/\iterationIndex$ \hfill (update upper search bound)
			
			\If{$\poMin<0$} 
			{
				$\poMin\gets0$ \hfill (ensure $\poMin\geq0$)
			}
			\If{$\poMax>2\pi$} 
			{
				$\poMax\gets2\pi$ \hfill (ensure $\poMax\leq2\pi$)
			}
			
		}
		
		\Return{$\poEst$}
		
	\end{algorithm}
	
	\subsection{Proposed Jutted-Zero Constellation} \label{subsec:constellation}
	
\begin{figure*}[t!]
			\centering
			
			\subfloat[Zero constellation.]{\includegraphics[width=1.75in]{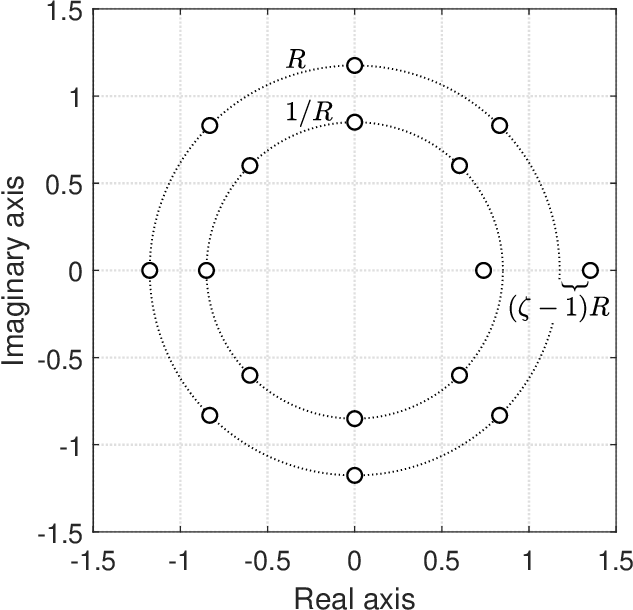}\label{subfig:jbmocz_constellation}}~
			\subfloat[Transmitted zeros.]{\includegraphics[width=1.75in]{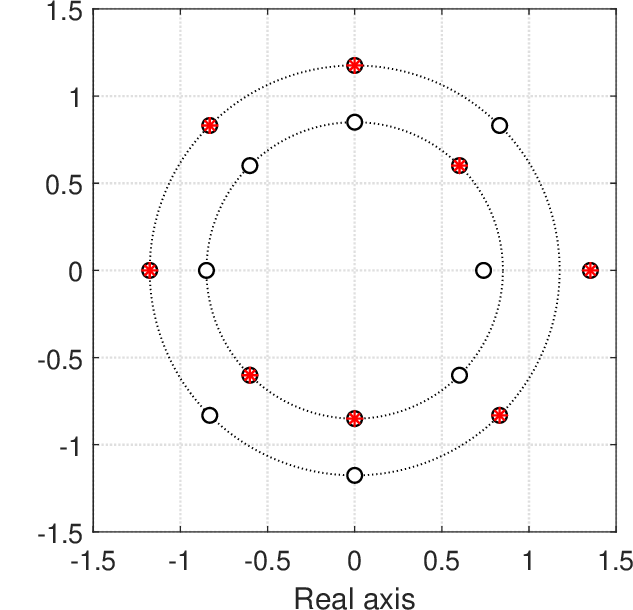}\label{subfig:jbmocz_tx}}~
			\subfloat[Received zeros.]{\includegraphics[width=1.75in]{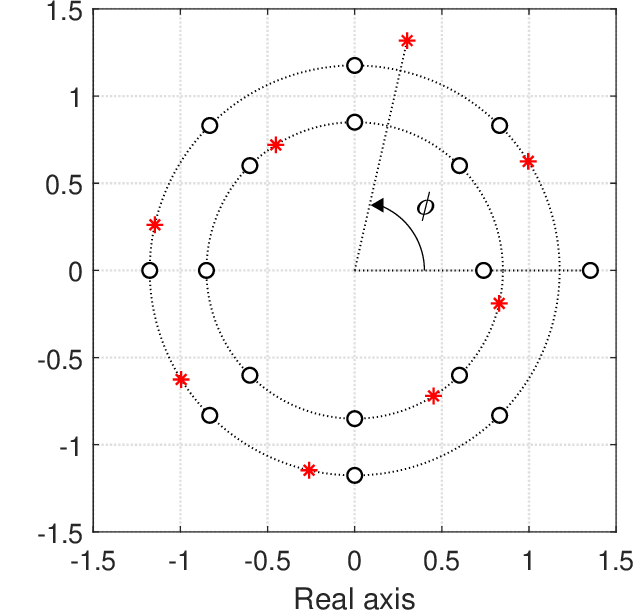}\label{subfig:jbmocz_rx}}~
			\subfloat[Corrected zeros.]{\includegraphics[width=1.75in]{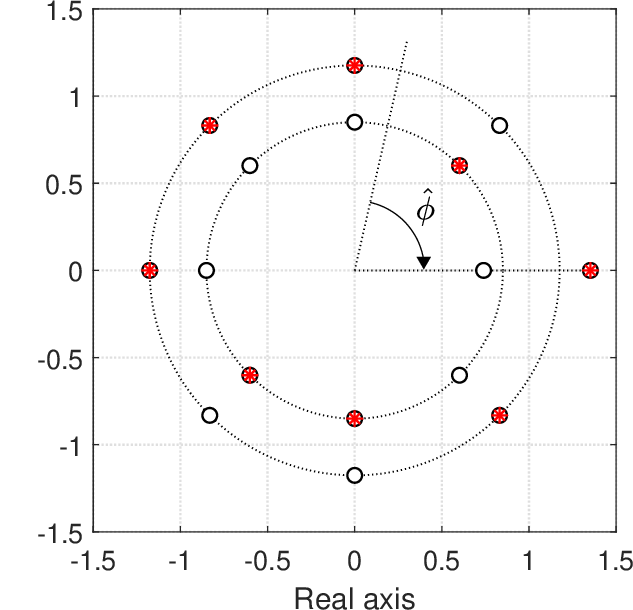}\label{subfig:jbmocz_corrected}}
			
			\caption{Proposed \ac{JBMOCZ} zero pattern for $\numZeros=8$, $\radius=1.176$, and $\asymFactor=1.15$. (a) Full zero constellation with $2\numZeros$ zero positions. (b) Transmitted zeros corresponding to the message $\bMessage=(1,0,1,1,1,0,0,1)$. (c) Received zeros rotated by $\po=(12/7)\baseAngle$ radians due to a \ac{CFO}. Received zeros after correcting $\po$ using Algorithm~\ref{alg:ICFOE}, which exactly correspond to the transmitted message.}	
			\label{fig:jbmocz_rotation}
	\end{figure*}

	To break the rotational symmetry of the Huffman \ac{BMOCZ} zero constellation while maintaining uniform phase separation of the zeros, we set $\zeroPhase_\zeroIndex\triangleq2\pi\zeroIndex/\numZeros$ and define
	\begin{equation} \label{eq:abmocz_mags}
		\zeroMag_\zeroIndex \triangleq
		\begin{cases}
			\asymFactor\radius, & \zeroIndex=0\\
			\radius, & \text{otherwise}
		\end{cases}\:, \ \ \zeroIndex\in[\numZeros]\:,
	\end{equation} 
	where $\asymFactor\geq1$ is an \emph{asymmetry factor}. With~\eqref{eq:abmocz_mags}, each of the zeros in~\eqref{eq:zero_mapping} lie on a circle of radius $\radius$ or $\radius^{-1}$, except for the zero corresponding to the first message bit, which is real and lies on a circle of radius $\asymFactor\radius\geq\radius$ or $(\asymFactor\radius)^{-1}\leq\radius^{-1}$. The asymmetry factor $\asymFactor$ controls the extent to which the first zero (i.e., $\zero_0$) ``juts out." When $\asymFactor=1$, the proposed zero constellation reduces to Huffman \ac{BMOCZ}. For $\asymFactor>1$, we call the approach \emph{jutted \ac{BMOCZ} (J-BMOCZ)}, an illustration of which is provided in Fig.~\ref{fig:jbmocz_rotation}. 
	
\begin{figure}[t!]
		\centering
		\includegraphics[width=3in]{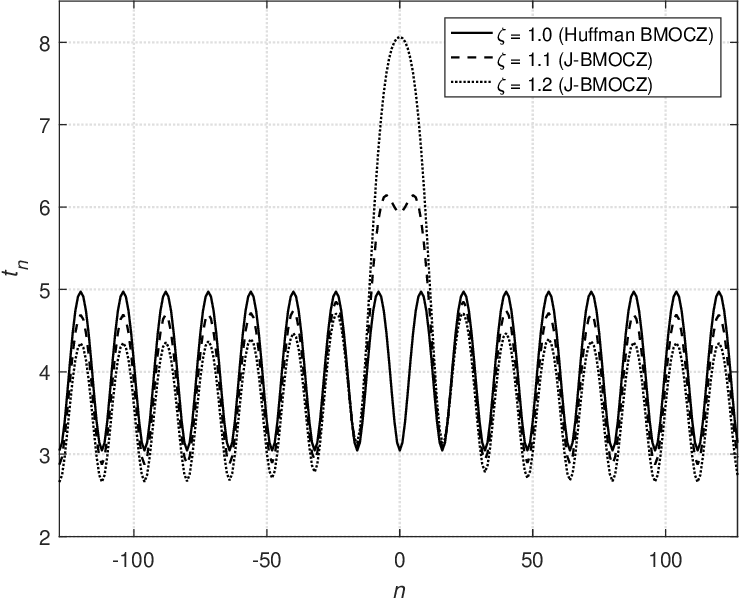}
		\caption{\ac{JBMOCZ} template vector for $\numZeros=16$, $\radius=1.093$, and $\dftSize=256$. Note: here $\dftIndex\in\{-128,\hdots,127\}$.}	
		\label{fig:template_plot}
	\end{figure}

	Fig.~\ref{fig:template_plot} shows the template vector $\tN$ for \ac{J-BMOCZ} with $\numZeros=16$ and $\asymFactor\in\{1,1.1,1.2\}$. Notice that for $\asymFactor>1$, a peak is introduced to $\tN$ that breaks the periodicity of the sinusoidal Huffman \ac{BMOCZ} template vector. The \ac{J-BMOCZ} template vector is thus well-suited for \ac{CFO} estimation using the approach of Section~\ref{subsec:fourier_approach}, as the inner product in~\eqref{eq:offsetEst} is maximized when the peak of $\tN$ aligns with the peak of a column in $\YNtilde$. We illustrate this concept with the following simple example.
	
	\begin{example} \label{example:walkthrough}
		Consider \ac{J-BMOCZ} with $\numZeros=2$, $\radius=1.5$, and $\asymFactor=1.2$. For $\dftSize=4$, the template vector becomes $\tN=[1.20,2.22,0.84,2.22]\trans$. Assume that we wish to transmit $\bMessage=(1,0)$. From~\eqref{eq:zero_mapping}, we get $\zero_0=\zeta\radius=1.8$ and $\zero_1=-1/\radius=-0.67$, yielding to the normalized transmit sequence $\txSeq=[1.08,1.02,-0.90]\trans$. For $h=\sqrt{0.5}(0.6+j)$ and $\po=\pi$, the received sequence $\rxSeqMod$ is given by~\eqref{eq:modulation} as
		\begin{align*}
			\rxSeqMod&=\modSeq_\po \odot \txSeq h \notag\\
			&=[0.46+0.76j,-0.43-0.72j,-0.38-0.63j]\trans\:,
  		\end{align*}
  		where we have omitted \ac{AWGN} for simplicity. Since $\dftSize=4$, with a single iteration of Algorithm~\ref{alg:ICFOE}, we can search over the rotations in $\{0,\pi/2,\pi,3\pi/2\}$. Computing $\YNtilde\in\C^{4\times4}$ with~\eqref{eq:dftAbsRX}, the inner product in~\eqref{eq:offsetEst} gives 
  		\begin{align*}
  			\tN\trans\YNtilde=[9.79,7.45,9.90,7.45]\:,
  		\end{align*}
  		leading to the estimate $\dftIndexHat=2\implies\poEst=(2\pi-0)\dftIndexHat/\dftSize=\pi$. Modulating $\rxSeqMod$ back with $\modSeq_{-\poEst}$, we obtain
  		\begin{align*}
  			\rxSeqHat&=\modSeq_{-\poEst} \odot \rxSeqMod \notag\\
  			&=[0.46+0.76j,0.43+0.72j,-0.38-0.63j]\trans\:.
  		\end{align*}
  		Finally, decoding $\Yhat(z)=\yhatelem_2z^2+\yhatelem_1z+\yhatelem_0$ via~\eqref{eq:dizet} yields
  		\begin{align*}
  			&|\Yhat(\asymFactor\radius)|=0.00<(\asymFactor\radius)^2|\Yhat((\asymFactor\radius)^{-1})|=2.04\implies \messageBit_0=1\:,\\
  			&|\Yhat(-\radius)|=4.38>\radius^2|\Yhat(-\radius^{-1})|=0.00\implies\messageBit_1=0\:,
  		\end{align*}
  		and hence $\hat{\bMessage}=(1,0)$. This concludes our example.
	\end{example}
	
	\section{Numerical Results} \label{sec:results}
	
	This section presents \ac{BER} and \ac{BLER} performance simulated as a function of $\EbNo$. In each simulation, we uniformly sample $\numBits$-bit messages $\bMessage\in\bField^\numBits$ and normalize $||\txSeq||_2^2$ to $\numZeros+1$. The transmitted polynomials are perturbed by \ac{AWGN} drawn from $\complexnormal(0,N_0)$, and in the fading channel, each polynomial observes a distinct channel gain $h\sim\complexnormal(0,1)$. To simulate a \ac{CFO}, each received sequence $\rxSeq\in\C^{\numZeros+1}$ is phase modulated by a unique angle $\po$ sampled uniformly from $[0,2\pi)$. Both Huffman \ac{BMOCZ} and \ac{JBMOCZ} employ the radius $\radius=\sqrt{1+\sin(\pi/\numZeros)}$ proposed in~\cite{walk2019principles}. For \ac{JBMOCZ}, we choose the asymmetry factor $\asymFactor$ by sweeping its value and selecting that which minimizes the \ac{BER} under a \ac{CFO}. All simulations use $\dftSize=64>\numZeros$, $\windowSize=0.2$, $\numIterations=2$, and the \ac{dizet} decoder in~\eqref{eq:dizet}.

	\subsection{Performance Loss Under \ac{CFO}} 
	
\begin{figure}[t!]
		\centering
		
		\subfloat[Uncoded bit error rate performance.]{\includegraphics[width=3in]{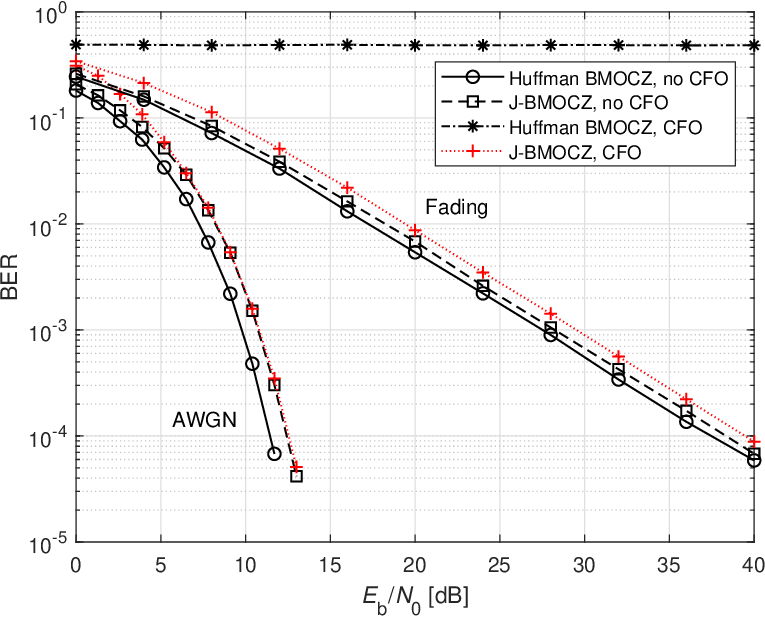}\label{subfig:ber_loss}}~\\
		\subfloat[Uncoded block error rate performance.]{\includegraphics[width=3in]{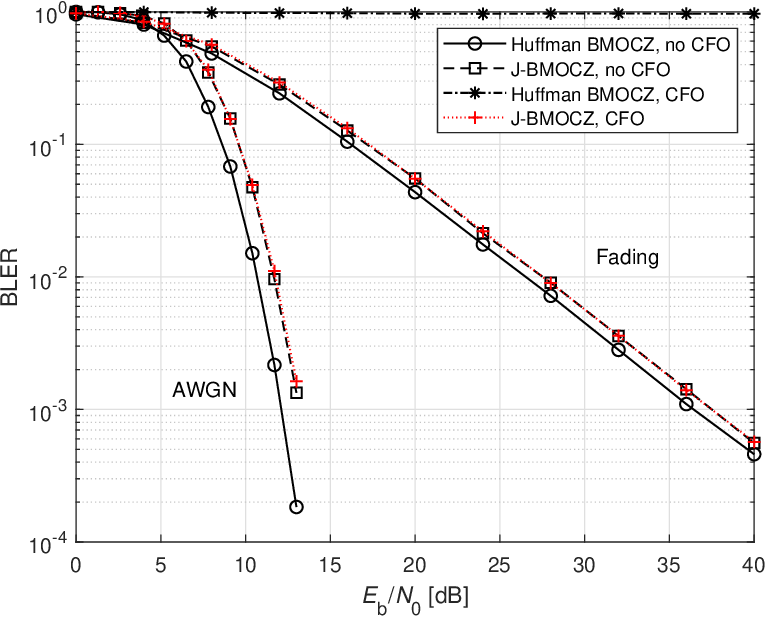}\label{subfig:bler_loss}}
		
		\caption{Comparison of uncoded \ac{JBMOCZ} to uncoded Huffman \ac{BMOCZ}. Both schemes use $\numZeros=32$ and $\radius=1.048$. The asymmetry factor for \ac{JBMOCZ} is $\asymFactor=1.15$.}	
		\label{fig:performance_loss}
	\end{figure}

	Fig.~\ref{fig:performance_loss} compares the performance of uncoded \ac{JBMOCZ} to uncoded Huffman \ac{BMOCZ}. In \ac{AWGN} without a \ac{CFO}, \ac{JBMOCZ} exhibits a 1~dB performance loss in both \ac{BER} and \ac{BLER} compared to Huffman \ac{BMOCZ}. In \ac{AWGN} under a \ac{CFO}, the \ac{BER} loss of \ac{JBMOCZ} relative to Huffman \ac{BMOCZ} without a \ac{CFO} remains roughly 1~dB, except at moderately low $\EbNo$, where the loss is slightly greater. In the fading channel without a \ac{CFO}, \ac{JBMOCZ} incurs a 0.7~dB loss in \ac{BER} and a 1~dB loss in \ac{BLER} compared to Huffman \ac{BMOCZ}. However, when a \ac{CFO} is introduced, the performance loss increases to 2~dB relative to Huffman \ac{BMOCZ} without a \ac{CFO}. These results are in contrast to the performance of uncoded Huffman \ac{BMOCZ} under a \ac{CFO}, since without a \ac{CPC}, the scheme cannot perform reliably and quickly encounters an error floor (see the black stars in Fig.~\ref{fig:performance_loss}).     
	
	\subsection{Comparison to Existing Methods}
	
	\begin{figure}[t]
		\centering
		
		\subfloat[Coded bit error rate performance.]{\includegraphics[width=3in]{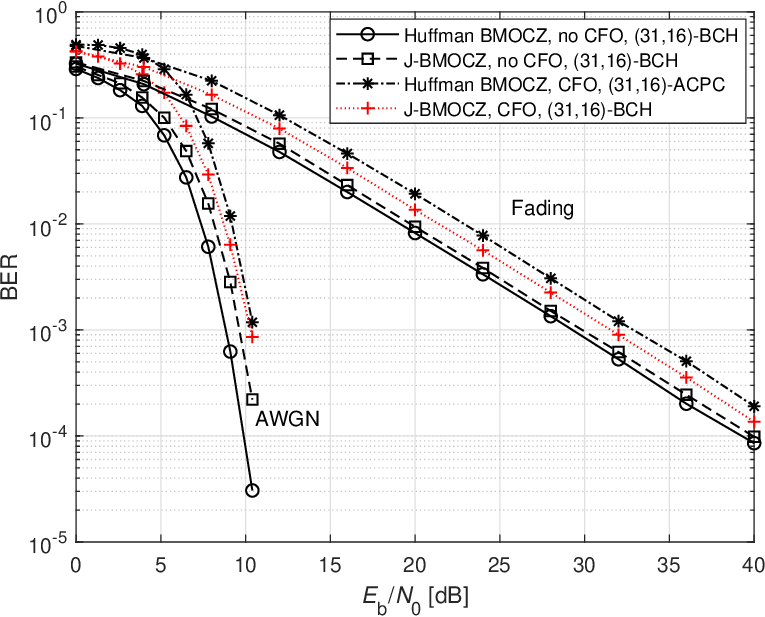}\label{subfig:coded_ber}}~\\
		\subfloat[Coded block error rate performance.]{\includegraphics[width=3in]{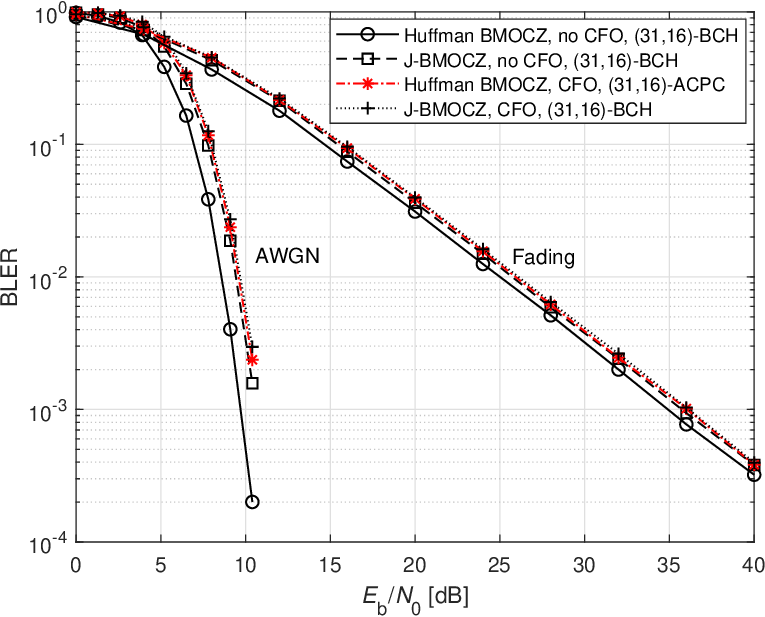}\label{subfig:coded_bler}}
		
		\caption{Comparison of coded \ac{JBMOCZ} to coded Huffman \ac{BMOCZ}. Both schemes use $\numZeros=31$ and $\radius=1.049$ to transmit $B=16$~bits. The asymmetry factor for \ac{JBMOCZ} is $\asymFactor=1.15$.}	
		\label{fig:coded_performance}
	\end{figure}
	
	Here, we compare the performance of coded \ac{JBMOCZ} to coded Huffman \ac{BMOCZ}. For Huffman \ac{BMOCZ}, we correct the fractional component of the \ac{CFO}, $\fracCFO$, using an oversampled \ac{IDFT}-based \ac{dizet} decoder with an oversampling factor of $Q=200$. To correct the integer component, $\intCFO$, we employ an \ac{ACPC} with a message length of $\numBits=16$~bits and a codeword length of $\numZeros=31$~bits. This (31,16)-\ac{ACPC} inherits a 2-bit error-correction capability from its outer (31,21)-BCH code. For further details on the \ac{IDFT}-based \ac{dizet} decoder and the (31,16)-\ac{ACPC} construction and implementation, we refer the reader to~\cite{walk2020practical}. For a fair comparison, we implement \ac{JBMOCZ} with a (31,16)-BCH code, which provides 3-bit error-correction capabilities at the same rate as the \ac{ACPC}. 
	
	Fig.~\ref{fig:coded_performance} shows the performance of coded \ac{JBMOCZ} and coded Huffman \ac{BMOCZ}. Similar to Fig.~\ref{fig:performance_loss}, Huffman \ac{BMOCZ} outperforms \ac{JBMOCZ} by 1~dB in \ac{AWGN} without a \ac{CFO}. However, in \ac{AWGN} with a \ac{CFO}, \ac{JBMOCZ}-BCH achieves better \ac{BER} performance than \ac{BMOCZ}-\ac{ACPC} at low $\EbNo$. In the fading channel, the \ac{BER} and \ac{BLER} losses of \ac{JBMOCZ} over Huffman \ac{BMOCZ} are reduced to just 0.5~dB and 1~dB, respectively. When a \ac{CFO} is introduced, \ac{JBMOCZ}-BCH outperforms \ac{BMOCZ}-\ac{ACPC} in \ac{BER} by roughly 1.75~dB while achieving similar \ac{BLER} performance. 
	
	\begin{remark}
		Although the (31,16)-BCH code improves the performance of both Huffman \ac{BMOCZ} and \ac{JBMOCZ} in \ac{AWGN}, it provides minimal benefit in the fading channel. Therefore, we suspect that a soft-decision implementation of the \ac{dizet} decoder, together with polar or \ac{LDPC} codes, could significantly improve the performance \ac{BMOCZ}. However, in the presence of a \ac{CFO}, such a decoder \emph{cannot} be implemented with Huffman \ac{BMOCZ}, since the \ac{ACPC} is required for \ac{CFO} correction. In contrast, \ac{JBMOCZ} enables decoding using soft information, as coding is not required to estimate the \ac{CFO}, and there are no restrictions on the message length. 
	\end{remark}
	
	\section{Conclusion} \label{sec:conclusion}
	
	In this work, we propose \ac{JBMOCZ}, a Huffman-like but asymmetric zero constellation featuring a ``jutted" zero pair. Furthermore, we exploit the common \ac{AACF} of \ac{BMOCZ} sequences to introduce an iterative \ac{CFO} estimation algorithm based on the \ac{IDFT}. In contrast to the existing methods for Huffman \ac{BMOCZ}, the proposed method for \ac{CFO} estimation with \ac{JBMOCZ} does not require pilots or channel coding. Using numerical simulations, we demonstrate the efficacy of \ac{JBMOCZ} under a \ac{CFO} by comparing its performance to Huffman \ac{BMOCZ} in both \ac{AWGN} and fading channels. Our results show that the performance loss of \ac{JBMOCZ} under a \ac{CFO} compared to Huffman \ac{BMOCZ} without a \ac{CFO} is minimal. Furthermore, we show that the \ac{BER} performance of coded \ac{JBMOCZ} outperforms Huffman \ac{BMOCZ} implemented with an \ac{ACPC}. Therefore, \ac{JBMOCZ} offers a robust and flexible alternative to Huffman \ac{BMOCZ} in scenarios where the use of a \ac{CPC} is impractical. In future work, we will focus on optimizing the proposed \ac{JBMOCZ} zero constellation, e.g., deriving the optimal $\asymFactor$ as a function of $\numZeros$. 
	
	\bibliographystyle{IEEEtran}
	\bibliography{references}
	
\end{document}